%% file: main.tex
\documentclass[10pt,conference,letterpaper]{IEEEtran}
\usepackage{amsmath,amsfonts}
\usepackage{algorithmic}
\usepackage{algorithm}
\usepackage{array}
\usepackage[caption=false,font=footnotesize,labelfont=sf,textfont=sf]{subfig}
\usepackage{textcomp}
\usepackage{stfloats}
\usepackage{url}
\usepackage{verbatim}
\usepackage{graphicx}
\usepackage{cite}
\usepackage{bm}
\usepackage[table]{xcolor}
\usepackage{xspace}

\newcommand{\new}[1]{#1}
\newcommand{\fss}{MPC$_\text{FSS}$\xspace}
\newcommand{\mpc}{MPC$_\textsc{A2B}$\xspace}
\newcommand{\fhe}{FHE\xspace}

\newcommand{\wan}{WAN$_\text{S}$\xspace}
\newcommand{\mwan}{WAN$_\text{M}$\xspace}
\newcommand{\fwan}{WAN$_\text{F}$\xspace}
\newcommand{\lan}{LAN$_\text{S}$\xspace}
\newcommand{\flan}{LAN$_\text{F}$\xspace}

\definecolor{heat1}{RGB}{255,255,204} 
\definecolor{heat2}{RGB}{255,237,160}
\definecolor{heat3}{RGB}{254,178,76}
\definecolor{heat4}{RGB}{253,141,60}
\definecolor{heat5}{RGB}{252,80,40}   

\usepackage{hyperref}
\usepackage{cleveref}
\usepackage{enumitem}

\usepackage{tikz}
\usetikzlibrary{arrows.meta}
\usetikzlibrary{calc}
\usetikzlibrary{shapes}
\usetikzlibrary{decorations.pathmorphing}
\usepackage{multirow}
\bstctlcite{BSTcontrol}
\begin{document}
\title{Beyond Latency: A System-Level Characterization of MPC and FHE for PPML}

\markboth{IEEE Journal of \LaTeX\ Class,~Vol.~12, No.~6, February~2024}%
{Shell \MakeLowercase{\textit{et al.}}: A Sample Article Using IEEEtran.cls for IEEE Journals}

\author{\IEEEauthorblockN{Pengzhi Huang}
\IEEEauthorblockA{
\textit{Cornell University}\\
Ithaca, NY, USA \\
ph448@cornell.edu}
\and
\IEEEauthorblockN{Kiwan Maeng}
\IEEEauthorblockA{
\textit{Pennsylvania State University}\\
University Park, PA, USA \\
kvm6242@psu.edu}
\and
\IEEEauthorblockN{G. Edward Suh}
\IEEEauthorblockA{
\textit{NVIDIA / Cornell University} \\
Westford, MA, USA \\
esuh@nvidia.com}
}
\IEEEpubid{0000-0000~\copyright~2024 IEEE}

\maketitle
\begingroup
\renewcommand\thefootnote{}
\footnotetext{This version includes minor corrections to the ISPASS 2026 accepted paper: a rounding inconsistency in Table IV and a coloring error in Figure 7. These changes do not affect the results or conclusions.}
\endgroup
\begin{abstract}
Privacy protection has become an increasing concern in modern machine learning applications.
Privacy-preserving machine learning (PPML) has attracted growing research attention, with approaches such as secure multiparty computation (MPC) and fully homomorphic encryption (FHE) being actively explored. 
However, existing evaluations of these approaches have frequently been done on a narrow, fragmented setup and have only focused on a specific performance metric, such as the online inference latency of a specific batch size.
From the existing reports, it is hard to compare different approaches, especially when considering other metrics like energy/cost or broader system setups (various hyperparameters, offline overheads, future hardware/network configurations, etc.).

We present a unified 
characterization of three popular approaches---two variants of MPC based on arithmetic/binary sharing conversion and function secret sharing, and FHE---on their performance and cost in performing privacy-preserving inference on multiple CNN and Transformer models.
We study a range of LAN and WAN environments, model sizes, batch sizes, and input sequence lengths. We evaluate not only the performance but also the energy consumption and monetary cost of deployment under a realistic scenario, taking into account their offline and online computation/communication overheads.
We provide empirical guidance for selecting, optimizing, and deploying these privacy-preserving compute paradigms, and outline how evolving hardware and network trends are likely to shift trade-offs between the two MPC schemes and FHE. This work provides system-level insights for researchers and practitioners who seek to understand or accelerate PPML workloads.
\end{abstract}


\section{Introduction}

Machine learning (ML) is now central to various applications ranging from healthcare diagnostics~\cite{kaissis2021end, wong2020deep, yang2017big} to cloud-based analytics services~\cite{Azure,GoogleAI}. These applications often need to process sensitive or proprietary data, such as patient records, corporate models, or personal identifiers, where confidentiality remains paramount. Cryptographic approaches for privacy-preserving machine learning (PPML) enable inference or training while protecting data and/or model privacy, thus offering stronger guarantees than regulatory or policy safeguards alone.

Two of the most studied paradigms are \emph{Secure Multiparty Computation (MPC)}~\cite{yao1982protocols,mohassel2017secureml,wagh2020falcon,jawalkar2024orca} and \emph{Fully Homomorphic Encryption (FHE)}~\cite{cheon2017homomorphic,kim2024cheddar,park2023toward,shivdikar2023gme}. MPC enables multiple parties to jointly evaluate a function over private inputs without revealing them to one another or to external observers. For ML workloads, MPC uses secret-sharing and interactive protocols to compute linear and non-linear layers of neural networks. MPC requires communication between participants and is sensitive to communication latency and bandwidth. Practical deployments must balance security, performance, and usability~\cite{zhou2024secure}.
FHE allows computation directly on encrypted data and does not rely on communication between parties during inference. Although relatively free from communication overhead, FHE requires orders of magnitude more computation compared to plaintext computation or MPC~\cite{zhang2024secure, kim2024cheddar}. 


In this paper, we focus on two popular variants of MPC: one that uses a combination of arithmetic and binary secret sharing (\mpc), and one that additionally adopts function secret sharing (\fss).
\mpc uses arithmetic sharing for linear layers and converts to binary sharing to evaluate major nonlinear operations. The approach has been popular in the community and is included in many frameworks~\cite{crypten2020, tan2021cryptgpu, wagh2020falcon, ma2023secretflow, mohassel2017secureml, kumar2020cryptflow, wagh2019securenn, dalskov2021fantastic, patra2020blaze}.
%
%
%
\fss, on the other hand, uses function secret sharing (FSS) for nonlinear operations~\cite{boyle2016function,de2022lightweight, gupta2023sigma, jawalkar2024orca}. \fss allows for an extensive offline phase (preprocessing) that generates keys for function-shares, enabling subsequent online evaluation with less communication overhead. 
\fss is relatively new, and studies on its system-level implications remain limited compared to \mpc. For FHE, we focus on \emph{CKKS}~\cite{cheon2017homomorphic}, which supports efficient vectorized approximate arithmetic on real-valued data and is therefore well suited for modern ML workloads. Most of recent FHE-based ML frameworks use CKKS~\cite{kim2024cheddar,zhang2024secure,park2023toward,shivdikar2023gme}.
In an FHE-based inference setting, the client encrypts its input and sends the ciphertext to the server, which performs all computations directly homomorphically with the ciphertext and returns the encrypted result to the client.




Because MPC and FHE adopt fundamentally different design philosophies---MPC heavily relies on communication, while FHE mostly relies on computation---their trade-offs span a rich multi-dimensional space. 
However, existing studies are limited in providing sufficient context for exploring and comparing these approaches. Many studies measure and report their performance on different setups~\cite{chen2025privacy,allaart2024private} and apply different optimizations to the model~\cite{zhang2024secure, kim2024cheddar}, making comparisons across different approaches difficult.
Also, many studies only report latency and throughput~\cite{cabrero2021sok, kim2024cheddar}, but not other deployment-related important metrics such as memory consumption, energy, and monetary cost, and focus on limited system/model configurations, making it difficult to generalize the results.



These observations motivate a holistic characterization study of \mpc, \fss, and \fhe under comparable setups across a range of system/model configurations. 
Our study compares four state-of-the-art frameworks~\cite{zhang2024secure, kim2024cheddar, liu2025depth, gupta2023sigma} that represent these PPML paradigms, reporting how various metrics (offline/online latency, throughput, energy consumption, cost of deployment, memory/storage overheads) change across different conditions and configurations, and project how they will scale with future hardware/network improvements.
%
Our key contributions are:

\begin{itemize}[leftmargin=*]
\item \textbf{A unified, apples-to-apples evaluation across three PPML paradigms.}  
We benchmark \mpc, \fss, and \fhe under the same hardware, network conditions, and CNN/Transformer workloads.  
This unified setup enables direct comparison of end-to-end latency, communication volume, and computation cost.
Our results provide a more controlled, consistent comparison of multiple encrypted computation techniques compared to previous studies that use different setups for different schemes.

\item \textbf{End-to-end system evaluation considering offline overhead, energy consumption, monetary cost, and memory/storage overhead.}  
Our analysis uncovers system-level behaviors and constraints that remain invisible when PPML is evaluated solely through latency and throughput:
(i) \fss is dominated by multi-gigabyte {offline} key-generation/distribution, which cannot be hidden/amortized in terms of monetary and energy cost;  
(ii) \mpc wastes significant CPU/GPU energy during its frequent online communication, revealing new optimization opportunities;  
(iii) The monetary cost of MPC is driven by total communication volume rather than communication latency or bandwidth under current volume-based pricing models in cloud networking services;
and 
(iv) \fhe incurs tens of GB of GPU memory pressure, while \fss requires large DRAM/SSD pools for key storage, creating limits on deployability.  
These observations show the importance of considering multi-dimensional aspects like energy, memory, and deployment cost when evaluating a PPML approach, not just latency.


\item \textbf{Scalability analysis across model size, batch size, context length, and hypothetical hardware scaling trend.}  
We show that \fhe scales worse than \mpc as activations or Transformer input contexts increase, due to compute-intensive ciphertext operations and high GPU memory usage.  
The scaling \mpc mainly depends on the activation dimension and communication volume.  
For {\fss}, its offline load grows rapidly with model size and query rate.  
Hardware scaling will likely favor \fhe as computation typically improves faster than communication. However, \fss still shows the lowest online latency among evaluated schemes if bandwidth and storage are sufficiently provisioned.  
These findings together highlight the need for joint protocol–model co-design to enable future large-scale PPML workloads.

\end{itemize}

\section{Background}

In this section, we briefly introduce how MPC and FHE work, focusing on the three schemes, \mpc, \fss, and CKKS \fhe.
While we mainly focus on their system characteristics, it should be noted that they have other trade-offs (in their threat models, supported use cases, etc.).

\subsection{Secure Multiparty Computation (MPC)}

In this paper, we consider MPC between two parties, where a sensitive value $x$ is split into two
secret shares held by two non-colluding parties, $P_1$ and $P_2$. 
A common secret sharing scheme is \emph{arithmetic sharing}, which splits the secret into two shares that add up to the original secret on an integer ring. That is, for a secret $x \in \mathbb{Z}/Q\mathbb{Z}$ and secret shares $[x]_1, [x]_2 \in \mathbb{Z}/Q\mathbb{Z}$ on an integer ring of size $2^Q$, 
\[
[x]_1 + [x]_2 \equiv x \ (\mathrm{mod}\ {Q}).
\]
Secret shares are distributed to each party ($P_1$ gets $[x]_1$ and $P_2$ gets $[x]_2$), and each share
looks random and reveals nothing about $x$ on its own.
Arithmetic sharing is commonly used in ML because many arithmetic computations can be done with these shares. For example,
to compute $z = x + y$, each party simply adds its own shares locally,
$[z]_i = [x]_i + [y]_i$, and it is easy to see that the result ($[z]_i$) is a secret share of the computation result ($z$). 

Computing \textbf{multiplications} and 
\textbf{nonlinear operations}, such as ReLU or Softmax, requires additional communication between parties.
Due to this additional communication, these operations become the dominant bottleneck in MPC, especially when the network between the two parties is slow (e.g., WAN).
%
%
Performing these operations also often requires pre-computation and/or agreeing upon shared values, which we refer to as the \textbf{offline phase}.
The offline phase can be done anytime as long as it happens before the \textbf{online phase}, which is when the precomputed materials are used to complete the evaluation upon an inference request.

Multiplications are commonly done through a technique called Beaver Triples~\cite{beaver1991efficient}, which relies on a pre-generated (during the offline phase) set of triples.
Different MPC frameworks differ in how they realize other nonlinear operations. 
\textbf{\mpc} protocols evaluates nonlinear operations generally through converting arithmetic shares into binary shares (arithmetic-to-binary conversion, or A2B), where each bit of the secret is split into a ring of size $Q=2$. Then a series of binary bit manipulations, arithmetic operations, and A2B/B2A conversions are performed in between.
Several popular frameworks that have been studied over the years can be broadly categorized into \mpc, as they all rely heavily on A2B conversion during nonlinear operations~\cite{crypten2020,wagh2020falcon,tan2021cryptgpu,wagh2019securenn}.
\textbf{\fss} protocols~\cite{gupta2023sigma, jawalkar2024orca, ryffel2020ariann}, in contrast, relies on function secret sharing (FSS) for nonlinear layers. FSS allows a function 
$f(\cdot)$ to be split into multiple function shares $(f_1(\cdot),f_2(\cdot))$, such that each party locally evaluates its share on an input, and the results combine to the secret-shared output: with $[y]_1=f_1(x),[y]_2=f_2(x)$, we can have $y = [y]_1+[y]_2=f(x)$. This enables parties to compute certain nonlinear functions with minimal online interaction, shifting most cryptographic work into an offline key-generation phase.
%

Compared to \mpc, \fss involves significantly more computation and communication in the offline phase, where large FSS keys (to represent cryptographically secure $(f_1(\cdot),f_2(\cdot))$) are generated and distributed, and more computation during the online evaluation of nonlinear operations.
In return, online communication during nonlinear operations, which is the main bottleneck of \mpc, becomes lighter, especially in terms of the number of communication rounds.
FSS keys cannot be reused across rounds or multiple inferences, so they have to be generated in large numbers.
%
%
%
Previous work~\cite{gupta2023sigma,jawalkar2024orca} commonly focused only on the online overhead, assuming that \emph{enough} keys are pre-generated, transmitted, and stored somewhere.

\subsection{Fully Homomorphic Encryption (FHE)}  
In this work, we focus on the CKKS scheme~\cite{cheon2017homomorphic} due to its support for efficient vectorized linear algebra on real-valued data, which is required for modern ML workloads. 
%
%
Unlike MPC, FHE does not require interactive communication during inference.
The client encrypts its input and sends the ciphertext (ct) to the server, and the server performs all the operations needed for ML inference locally using homomorphic computation. In the end, the (encrypted) result is sent back to the client and can be decrypted by client's secret keys.
The weights remain in plaintext (pt) and are kept only on the server-side~\cite{kim2024cheddar,zhang2024secure}.

As FHE only requires a small amount of communication at the beginning and end, its performance does not significantly depend on the network speed. However, FHE is much more computation-heavy than MPC. FHE operations are more expensive than their plaintext counterparts, and as each operation injects noise, an expensive procedure called \emph{bootstrapping} must be done after a certain number of operations to reset the accumulated noise.
Several previous work~\cite{kim2024cheddar,park2023toward,shivdikar2023gme} reported that bootstrapping dominates the overhead when running ML inference.
FHE also incurs a larger memory footprint due to the ciphertext being much larger than the plaintext.
FHE can only perform addition or multiplication, so functions such as ReLU, max, or Sigmoid must be approximated through polynomials.
Similarly, operations like max pooling are often replaced with other operations like average pooling.

We focus on GPU-based implementations~\cite{kim2024cheddar, zhang2024secure} of FHE-based ML, which represents the cost of deploying an FHE-based system on existing devices.
Other work~\cite{kim2022bts, samardzic2021f1, samardzic2022craterlake} has proposed custom accelerators for FHE, which can significantly improve the performance and the cost of deploying an FHE-based system.
To the best of our knowledge, these accelerators have not been commercialized yet and cannot be used for deployment today.
There are hybrid systems that perform linear layers with FHE and nonlinear layers with MPC~\cite{reagen2021cheetah, juvekar2018gazelle, pang2024bolt, lu2023bumblebee, rathee2020cryptflow2, mishra2020delphi}.
%
We do not directly study these systems but expect them to show characteristics that lie somewhere between the two approaches (FHE and MPC). 

\section{Methodology}

To ensure fairness and reproducibility, we combine empirical measurements, analytical latency modeling, and derived energy and monetary estimates across all protocols. This hybrid approach allows for consistent evaluation despite differing public implementations.

\subsection{Experimental Setup}
\label{sec:experimental_setup}

\textbf{System setup.}
We ran all the experiments on nodes equipped with an Intel Xeon Gold 6448Y CPU, 512GB main memory, and an NVIDIA A6000 GPU (multiple nodes for \mpc/\fss runs).
%
All experiments
were run with the GPU clock fixed using persistence mode. CUDA 12.5-compatible NVIDIA drivers and PyTorch 2.6 with \texttt{cu121} were used. 

\textbf{Network.}
We studied various network setups, summarized in~\autoref{tab:setup}: three WAN setups with a Round Trip Time (RTT) latency of 70 ms and varying bandwidth, slow WAN (\wan, 70 MBps), medium WAN (\mwan, 1GBps), and fast WAN (\fwan, 50 GBps); and two LAN setups with RTT of 0.02ms, typical LAN (\lan, 1 GBps) and fast LAN (\flan, 50 GBps). The \mwan is chosen to match the \lan speed used in previous work~\cite{gupta2023sigma,jawalkar2024orca}, and the 50GBps speed is chosen to be comparable to the fastest offered by the existing commercial product (AWS Direct Connect~\cite{aws_direct_connect}). In practice, however, MPC requires multiple non-colluding parties, which is difficult to guarantee when all servers are physically located at the same place and even controlled by a single provider, making pure LAN deployments less realistic. Consequently, we treat WAN as the primary deployment regime and use LAN (mainly \lan) as an optimistic reference point. We emulate MPC performance on these networks by running two parties on a single node with separate resources and estimating communication costs using the measured number of rounds and packet sizes.
\begin{table}[t]
\centering
\caption{Experimental configuration for schemes and network. Both \fhe systems are CKKS-based and \fss systems use the same FSS framework~\cite{boyle2016function}.
}
\small
\begin{tabular}{l l l}
\hline
\textbf{Scheme} & \textbf{Protocol/System} & \\
\hline
\fhe &
\multicolumn{2}{l}{Cheddar~\cite{kim2024cheddar} for ResNet, NEXUS~\cite{zhang2024secure} for BERT} \\
\mpc &
\multicolumn{2}{l}{Optimized CrypTen~\cite{cryptorch, liu2025depth}} \\
\fss &
\multicolumn{2}{l}{Orca~\cite{jawalkar2024orca} for ResNet, Sigma~\cite{gupta2023sigma} for BERT} \\
\hline
\textbf{Network} & \textbf{Latency (RTT)} & \textbf{Bandwidth} \\
\hline
\wan & 70 ms & 70 MBps\\
\mwan & 70 ms & 1 GBps\\
\fwan & 70 ms & 50 GBps\\
\lan & 0.02 ms & 1 GBps\\
\flan & 0.02 ms & 50 GBps\\
\hline
\end{tabular}
\label{tab:setup}
\end{table}


\begin{table}[t]
\centering
\small
\caption{Model parameter counts and approximate computational complexity. FLOPs are per inference for the input sizes shown.}
\begin{tabular}{lcccc}
\hline
Model & Params & Layers &  Input & FLOPs \\
\hline
BERT-Tiny & 4.4M & 2  & 128 tokens & $\sim$0.4G \\
BERT-Base & 110M & 12 & 128 tokens & $\sim$22G  \\
ResNet-20 & 0.27M & 20 & $32\times32$  & $\sim$0.04G \\
ResNet-50 & 25.6M & 50 & $224\times224$ & $\sim$4.1G \\
\hline
\end{tabular}

\label{tab:model_size}
\end{table}

\textbf{Models.} We evaluated four models that are commonly studied in the literature~\cite{kim2024cheddar, jawalkar2024orca, gupta2023sigma, maeng2024accelerating, liu2025depth, zhang2024secure}: ResNet-20 with CIFAR-10, ResNet-50 with ImageNet, BERT-Tiny, and BERT-Base (\autoref{tab:model_size}). 
%

For BERT, we use a sequence length of 128 and vary it in some experiments.
For ResNet models, we replaced max pooling with average pooling, following previous work~\cite{kim2024cheddar, maeng2024accelerating}.
For the FHE evaluation of BERT, the max operator preceding Softmax (in FHE/MPC, a max operator typically precedes Softmax for numerical stability~\cite{crypten2020, gupta2023sigma}) is replaced with a constant value in NEXUS~\cite{zhang2024secure}. An exact max requires iterative pairwise comparisons using a large number of bootstrapping, which imposes prohibitively high overhead in practice. 
\new{Max only accounts for $<$10\%/$<$20\% of the batched/unbatched MPC overheads. Applying the same approximation used in FHE would therefore not significantly change the qualitative trends.} Other literature avoids this complexity by replacing softmax~\cite{chen2022x,zimerman2024converting}, using alternative normalization~\cite{moon2025thor}, or evaluating only on a reduced scale~\cite{rovida2024transformer}.  \new{The accuracy impact of the above modification is minimal, as documented by prior work (e.g., BERT-base: $0.20$--$0.40\%$ drop for FHE~\cite{zhang2024secure}, negligible for MPCs~\cite{gupta2023sigma}).}

%

When evaluating per-sample latency, we use a batch size of 1. When measuring throughput, we use a batch size of 128 for MPC, as batching significantly improves throughput by amortizing network round latency.
We always use a batch size of 1 for FHE, following previous work~\cite{kim2024cheddar}, because GPU memory is already heavily utilized by the ciphertext-level operations of a single sample inference. Memory becomes a major bottleneck, as discussed in~\autoref{sec:mem}.

\subsection{Evaluated Frameworks}

\textbf{FHE.}
\autoref{tab:setup} summarizes the libraries we used for each approach.
For FHE, we used Cheddar~\cite{kim2024cheddar} for ResNet and NEXUS~\cite{zhang2024secure} for BERT, which are the two state-of-the-art GPU FHE libraries (we used two CKKS-based libraries, as each can only handle CNNs and Transformers, respectively, but not both).
While newer work~\cite{fan2025warpdrive} claims improved performance, the code is not open-sourced yet, and we could not include it in the evaluation.
All experiments used 128-bit security, a polynomial degree of $N = 2^{16}$, and the recommended configurations of Cheddar/NEXUS.

As our setup uses a single GPU per node, some layers of models could not fit in the GPU memory.
We use extrapolated scaling to provide additional insight into what the results might look like for hypothetical setups with enough memory. For layers that are too large to run (the $\mathbb{R}^{128 \times 3072} \times \mathbb{R}^{3072 \times 768}$ layer of BERT-Base and several layers of ResNet-50), we estimated the execution time by linearly scaling the latency of a smaller layer with the computation complexity.
As a result, when the full model cannot fit on our GPU, the reported \fhe numbers should be interpreted as analytical runtime estimates rather than exact measurements. \new{We mark this difference in our results with $\ast$ symbols.}


%




\textbf{MPC.}
For \mpc, we used the optimized variant of CrypTen~\cite{crypten2020} published as part of the CrypTorch library~\cite{cryptorch, liu2025depth}.
%
For \fss, we used Orca~\cite{jawalkar2024orca} for ResNet and its follow-up work, Sigma~\cite{gupta2023sigma} for BERT.
%
\mpc and \fss incur separate offline (Beaver Triples or key generation and distribution) and online costs (online inference).
We separately measure and report these online/offline overheads.
The online-only overhead represents an ideal case when the offline compute is done during idle times with no queries, while adding both overheads (online+offline) represents a case where the offline cost cannot be hidden, e.g., when the query arrival rate is higher than the offline compute throughput.

\new{Since our characterization is based on current mainstream protocols and prevailing design trends, we do not treat embedding as a major bottleneck and assume tokenization is done on the client device before sending the query to the FHE/MPC server. This assumption is consistent with prior work~\cite{kim2024cheddar, jawalkar2024orca, gupta2023sigma,crypten2020,liu2025depth}. Workloads requiring large private embedding tables/private retrieval introduce an additional dimension of challenges beyond the scope of this study.}



\subsection{Energy and Monetary Cost Estimation}\label{sec:methoenergy}

\textbf{Energy.}
CPU and GPU power consumptions are measured via Intel RAPL~\cite{david2010rapl} and NVIDIA monitoring utilities, respectively.
For NIC power consumption, we use the value from
Mellanox ConnectX-5 MCX515A-CCAT 
specifications~\cite{ConnectX5-spec}. ~\autoref{tab:power} summarizes the collected power numbers. 
The CPU power consumption during \mpc computation is roughly the same as that of medium-speed transmission. \new{Power measurements were collected under representative computation and communication workloads, and total energy was estimated by multiplying phase-specific power with execution time. Although these are not direct end-to-end measurements, they provide reasonable approximations for comparative analysis.}

\begin{table}[t]
\centering
\caption{Collected power numbers (W)}
\label{tab:power}
\begin{tabular}{|c|c||c|c|c|c|}
\hline
\textbf{Comp.}&\textbf{GPU}  & \textbf{Comm. Conf.}& \textbf{CPU} & \textbf{MemIO}
 & \textbf{NIC/ea.} \\\hline
Heavy & 250  & High speed & 60  & 15 & 25 \\\hline
Light & 50    &  Med speed & 25 & 5 & 10\\\hline
Idle& 6 & Idle/Low speed & 16 & 3 & 5\\\hline
\end{tabular}
\end{table}

\textbf{Monetary cost.}
We used AWS pricing~\cite{aws-ec2-pricing,aws_direct_connect} to roughly estimate the monetary cost per inference, including the cost of the compute instance, storage, and network data transfer charges.
For the compute instance, we used the hourly cost of the AWS p3.2xlarge instance as the closest single-GPU service to our experimental setting and converted it to a per-second rate of roughly \$0.001/s.
%
For storage, we used \$9.6$\times$10$^{-5}$/s for a volume of 5~TB.
When using high-speed LAN/WAN (\fwan/\flan), AWS bills \$0.0236/s for port usage and \$0.02/GB for data transfer. 

%


\section{Experimental Results}
In this section, we present our results and summarize the key findings in~\autoref{sec:takeaways}.

\subsection{Latency and Throughput}

\begin{table*}[h!]
\centering
\caption{Per-sample latency (seconds; measured with a batch size of 1) across different network configurations. $\ast$ denotes estimated results.
}
\begingroup
\setlength{\tabcolsep}{3pt}
\begin{tabular}{|l@{}|cc|cc|c@{}|cc|cc|c@{}|cc|cc|c@{}|cc|cc|c|}
\hline
\multirow{3}{*}{\textbf{Setting}} &
\multicolumn{5}{c|}{\textbf{BERT-Tiny}} &
\multicolumn{5}{c|}{\textbf{BERT-Base}} &
\multicolumn{5}{c|}{\textbf{ResNet-20}} &
\multicolumn{5}{c|}{\textbf{ResNet-50}} \\ \cline{2-21}
&
\multicolumn{2}{c|}{MPC On+Off} & \multicolumn{2}{c|}{MPC Online} & \multirow{2}{*}{FHE}  &
\multicolumn{2}{c|}{MPC On+Off} & \multicolumn{2}{c|}{MPC Online} & \multirow{2}{*}{FHE*}  &
\multicolumn{2}{c|}{MPC On+Off} & \multicolumn{2}{c|}{MPC Online} & \multirow{2}{*}{FHE}  &
\multicolumn{2}{c|}{MPC On+Off} & \multicolumn{2}{c|}{MPC Online} & \multirow{2}{*}{FHE*}  \\ \cline{2-5} \cline{7-10} \cline{12-15} \cline{17-20}
 & FSS & A2B & FSS & A2B & 
 & FSS & A2B & FSS & A2B & 
 & FSS & A2B & FSS & A2B & 
 & FSS & A2B & FSS & A2B &  \\ \hline

\lan &
  \cellcolor{heat4}{0.84} & \cellcolor{heat2}{\underline{0.24}} &
  \cellcolor{heat3}{0.29} & \cellcolor{heat1}{\underline{0.23}} &
  \cellcolor{heat5}{4.0} &

  \cellcolor{heat4}{24}   & \cellcolor{heat3}{\underline{9.1}} &
  \cellcolor{heat1}{\underline{2.3}} & \cellcolor{heat2}{8.2} &
  \cellcolor{heat5}{149} &

  \cellcolor{heat4}{0.21}    & \cellcolor{heat3}{\underline{0.074}} &
  \cellcolor{heat1}{\underline{0.024}} & \cellcolor{heat2}{0.058} &
  \cellcolor{heat5}{1.7} &

  \cellcolor{heat4}{15} & \cellcolor{heat3}{\underline{2.5}} &
  \cellcolor{heat1}{\underline{1.0}} & \cellcolor{heat2}{2.2} &
  \cellcolor{heat5}{164}
\\\hline

\flan &
  \cellcolor{heat3}{0.39} & \cellcolor{heat1}{\underline{0.15}} &
  \cellcolor{heat2}{0.26} & \cellcolor{heat1}{\underline{0.15}} &
  \cellcolor{heat4}{4.0} &

  \cellcolor{heat4}{5.9} & \cellcolor{heat3}{\underline{5.3}} &
  \cellcolor{heat1}{\underline{1.3}} & \cellcolor{heat2}{5.2} &
  \cellcolor{heat5}{149} &

  \cellcolor{heat4}{0.053}   & \cellcolor{heat3}{\underline{0.034}} &
  \cellcolor{heat1}{\underline{0.017}} & \cellcolor{heat2}{0.033} &
  \cellcolor{heat5}{1.7} &

  \cellcolor{heat4}{5.4} & \cellcolor{heat3}{\underline{0.99}} &
  \cellcolor{heat1}{\underline{0.84}} & \cellcolor{heat2}{0.97} &
  \cellcolor{heat5}{164}
\\\hline

\wan &
  \cellcolor{heat3}{\underline{12}} & \cellcolor{heat5}{33} &
  \cellcolor{heat2}{\underline{6.0}} & \cellcolor{heat4}{33} &
  \cellcolor{heat1}{\underline{4.0}} &

  \cellcolor{heat5}{301} & \cellcolor{heat4}{\underline{243}} &
  \cellcolor{heat1}{\underline{47}} & \cellcolor{heat3}{232} &
  \cellcolor{heat2}{\underline{149}} &

  \cellcolor{heat3}{\underline{9.4}} & \cellcolor{heat5}{14} &
  \cellcolor{heat2}{\underline{7.0}} & \cellcolor{heat4}{14} &
  \cellcolor{heat1}{\underline{1.7}} &

  \cellcolor{heat5}{168} & \cellcolor{heat3}{\underline{59}} &
  \cellcolor{heat1}{\underline{18}} & \cellcolor{heat2}{55} &
  \cellcolor{heat4}{164}
\\\hline

\mwan &
  \cellcolor{heat3}{\underline{6.2}} & \cellcolor{heat5}{32} &
  \cellcolor{heat2}{\underline{5.6}} & \cellcolor{heat4}{32} &
  \cellcolor{heat1}{4.0} &

  \cellcolor{heat2}{\underline{55}} & \cellcolor{heat5}{187} &
  \cellcolor{heat1}{\underline{34}} & \cellcolor{heat4}{186} &
  \cellcolor{heat3}{149} &

  \cellcolor{heat3}{\underline{7.1}} & \cellcolor{heat5}{14} &
  \cellcolor{heat2}{\underline{6.9}} & \cellcolor{heat4}{14} &
  \cellcolor{heat1}{1.7} &

  \cellcolor{heat2}{\underline{29}} & \cellcolor{heat3}{37} &
  \cellcolor{heat1}{\underline{16}} & \cellcolor{heat3}{37} &
  \cellcolor{heat5}{164}
\\\hline

\fwan &
  \cellcolor{heat3}{\underline{5.7}} & \cellcolor{heat5}{32} &
  \cellcolor{heat2}{\underline{5.6}} & \cellcolor{heat4}{32} &
  \cellcolor{heat1}{\underline{4.0}} &

  \cellcolor{heat2}{\underline{38}} & \cellcolor{heat4}{183} &
  \cellcolor{heat1}{\underline{33}} & \cellcolor{heat5}{182} &
  \cellcolor{heat3}{149} &

  \cellcolor{heat3}{\underline{7.0}} & \cellcolor{heat5}{14} &
  \cellcolor{heat2}{\underline{6.9}} & \cellcolor{heat4}{14} &
  \cellcolor{heat1}{\underline{1.7}} &

  \cellcolor{heat2}{\underline{20}} & \cellcolor{heat3}{36} &
  \cellcolor{heat1}{\underline{16}} & \cellcolor{heat3}{36} &
  \cellcolor{heat5}{164}
\\\hline
\end{tabular}
\endgroup
\label{tab:latency}
\end{table*}

\begin{table*}[h!]
\centering
\caption{Throughput (samples/s;  measured with a batch size of 128 for MPC, 1 for FHE, as explained in ~\autoref{sec:experimental_setup}).}
\begingroup
\setlength{\tabcolsep}{3pt}
\begin{tabular}{|l@{}|cc|cc|c@{}|cc|cc|c@{}|cc|cc|c@{}|cc|cc|c|}
\hline
\multirow{3}{*}{\textbf{Setting}} &
\multicolumn{5}{c|}{\textbf{BERT-Tiny}} &
\multicolumn{5}{c|}{\textbf{BERT-Base}} &
\multicolumn{5}{c|}{\textbf{ResNet-20}} &
\multicolumn{5}{c|}{\textbf{ResNet-50}} \\ \cline{2-21}
&
\multicolumn{2}{c|}{MPC On+Off} & \multicolumn{2}{c|}{MPC Online} & \multirow{2}{*}{FHE}  &
\multicolumn{2}{c|}{MPC On+Off} & \multicolumn{2}{c|}{MPC Online} & \multirow{2}{*}{FHE*}  &
\multicolumn{2}{c|}{MPC On+Off} & \multicolumn{2}{c|}{MPC Online} & \multirow{2}{*}{FHE}  &
\multicolumn{2}{c|}{MPC On+Off} & \multicolumn{2}{c|}{MPC Online} & \multirow{2}{*}{FHE*}  \\ \cline{2-5} \cline{7-10} \cline{12-15} \cline{17-20}
 & FSS & A2B & FSS & A2B & 
 & FSS & A2B & FSS & A2B & 
 & FSS & A2B & FSS & A2B & 
 & FSS & A2B & FSS & A2B &  \\ \hline
\lan &
  \cellcolor{heat4}{1.6} & \cellcolor{heat3}{\underline{4.6}} &
  \cellcolor{heat1}{\underline{17}} & \cellcolor{heat2}{4.7} &
  \cellcolor{heat5}{0.25} &

  \cellcolor{heat4}{0.04} & \cellcolor{heat3}{\underline{0.17}} &
  \cellcolor{heat1}{\underline{0.50}} & \cellcolor{heat2}{0.20} &
  \cellcolor{heat5}{0.007} &

  \cellcolor{heat4}{4.9} & \cellcolor{heat3}{\underline{16}} &
  \cellcolor{heat1}{\underline{53}} & \cellcolor{heat2}{21} &
  \cellcolor{heat5}{0.60} &

  \cellcolor{heat4}{0.07} & \cellcolor{heat3}{\underline{0.40}} &
  \cellcolor{heat1}{\underline{0.99}} & \cellcolor{heat2}{0.45} &
  \cellcolor{heat5}{0.006}
\\\hline

\flan &
  \cellcolor{heat4}{6.3} & \cellcolor{heat3}{\underline{7.8}} &
  \cellcolor{heat1}{\underline{31}} & \cellcolor{heat2}{7.9} &
  \cellcolor{heat5}{0.25} &

  \cellcolor{heat4}{0.18} & \cellcolor{heat3}{\underline{0.45}} &
  \cellcolor{heat1}{\underline{0.93}} & \cellcolor{heat2}{0.49} &
  \cellcolor{heat5}{0.007} &

  \cellcolor{heat4}{24} & \cellcolor{heat3}{\underline{41}} &
  \cellcolor{heat1}{\underline{83}} & \cellcolor{heat2}{43} &
  \cellcolor{heat5}{0.60} &

  \cellcolor{heat4}{0.23} & \cellcolor{heat3}{\underline{1.0}} &
  \cellcolor{heat1}{\underline{1.2}} & \cellcolor{heat2}{1.1} &
  \cellcolor{heat5}{0.006}
\\\hline

\wan &
  \cellcolor{heat5}{0.15} & \cellcolor{heat3}{\underline{0.58}} &
  \cellcolor{heat1}{\underline{2.2}} & \cellcolor{heat2}{0.61} &
  \cellcolor{heat4}{0.25} &

  \cellcolor{heat5}{0.003} & \cellcolor{heat3}{\underline{0.017}} &
  \cellcolor{heat1}{\underline{0.07}} & \cellcolor{heat2}{0.02} &
  \cellcolor{heat4}{0.007} &

  \cellcolor{heat5}{0.40} & \cellcolor{heat3}{\underline{1.4}} &
  \cellcolor{heat1}{\underline{6.0}} & \cellcolor{heat2}{2.0} &
  \cellcolor{heat4}{0.60} &

  \cellcolor{heat4}{0.007} & \cellcolor{heat3}{\underline{0.04}} &
  \cellcolor{heat1}{\underline{0.28}} & \cellcolor{heat2}{0.05} &
  \cellcolor{heat5}{0.006}
\\\hline

\mwan &
  \cellcolor{heat4}{1.5} & \cellcolor{heat3}{\underline{2.1}} &
  \cellcolor{heat1}{\underline{10}} & \cellcolor{heat2}{2.2} &
  \cellcolor{heat5}{0.25} &

  \cellcolor{heat4}{0.042} & \cellcolor{heat3}{\underline{0.14}} &
  \cellcolor{heat1}{\underline{0.44}} & \cellcolor{heat2}{0.15} &
  \cellcolor{heat5}{0.007} &

  \cellcolor{heat4}{3.9} & \cellcolor{heat3}{\underline{5.9}} &
  \cellcolor{heat1}{\underline{13}} & \cellcolor{heat2}{6.4} &
  \cellcolor{heat5}{0.60} &

  \cellcolor{heat4}{0.070} & \cellcolor{heat3}{\underline{0.36}} &
  \cellcolor{heat1}{\underline{0.89}} & \cellcolor{heat2}{0.40} &
  \cellcolor{heat5}{0.006}
\\\hline

\fwan &
  \cellcolor{heat3}{\underline{5.0}} & \cellcolor{heat4}{2.7} &
  \cellcolor{heat1}{\underline{14}} & \cellcolor{heat4}{2.7} &
  \cellcolor{heat5}{0.25} &

  \cellcolor{heat4}{0.17} & \cellcolor{heat3}{\underline{0.28}} &
  \cellcolor{heat1}{\underline{0.75}} & \cellcolor{heat2}{0.29} &
  \cellcolor{heat5}{0.007} &

  \cellcolor{heat2}{\underline{10}} & \cellcolor{heat4}{7.6} &
  \cellcolor{heat1}{\underline{15}} & \cellcolor{heat3}{7.7} &
  \cellcolor{heat5}{0.60} &

  \cellcolor{heat4}{0.22} & \cellcolor{heat3}{\underline{0.81}} &
  \cellcolor{heat1}{\underline{1.1}} & \cellcolor{heat2}{0.82} &
  \cellcolor{heat5}{0.006}
\\\hline

\end{tabular}
\endgroup
\label{tab:throughput}
\end{table*}

\autoref{tab:latency} summarizes the estimated end-to-end latencies with a batch size of 1, and \autoref{tab:throughput} summarizes the throughput (batch size of 128 for \mpc/\fss, batch size of 1 for \fhe) under different network configurations. The underlined values show the best-performing \mpc/\fss protocol for online or offline; \fhe is underlined too when it outperforms both MPC variants. Colors indicate relative performance within each model/network setup combination, with deeper red indicating worse and lighter yellow indicating better performance.

\textbf{Overall trend.}
Under low-latency and/or high-bandwidth network configurations (\lan, \flan, and \fwan), MPC achieves the best end-to-end latency and throughput. \fss provides the lowest online latency but requires heavier offline computation. As RTT increases and bandwidth decreases (\wan, \mwan), \mpc performance degrades due to its higher round complexity and online communication volume. \fss is slightly less sensitive to the increases in RTT, but the huge volume of offline communication becomes the bottleneck when considering both offline and online costs. We also observe that within the same model family, \fhe incurs a larger increase in latency (decrease in throughput) as model size scales, compared to MPC protocols.


\begin{figure}[t]
    \centering
    \subfloat[BERT-Base\label{fig:latency_sub}]{
        \includegraphics[width=0.5\linewidth]{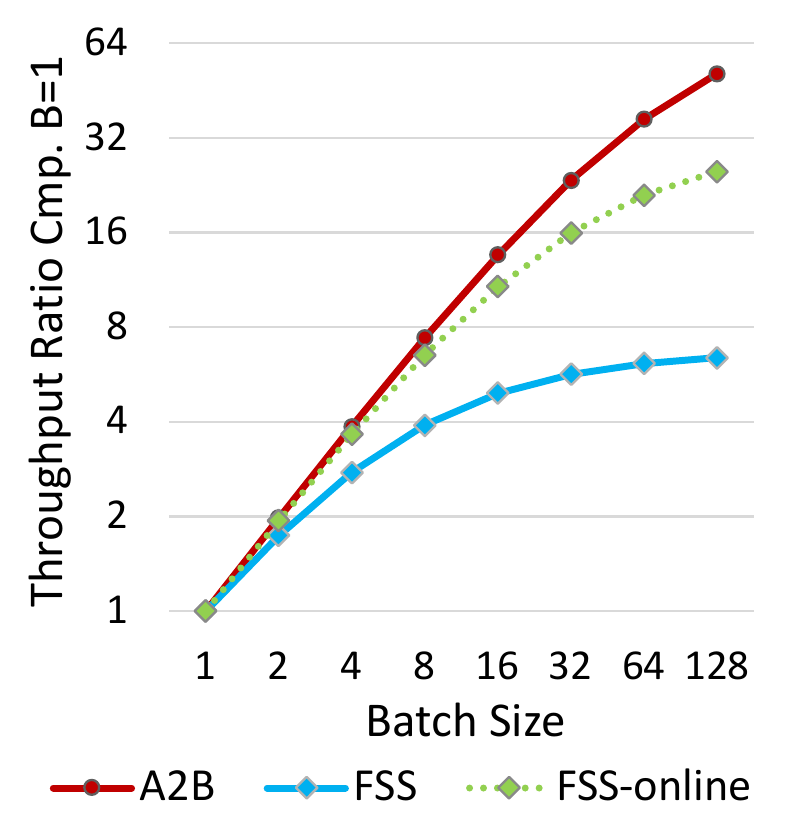}
    }
    \subfloat[ResNet-50\label{fig:throughput_sub}]{
        \includegraphics[width=0.5\linewidth]{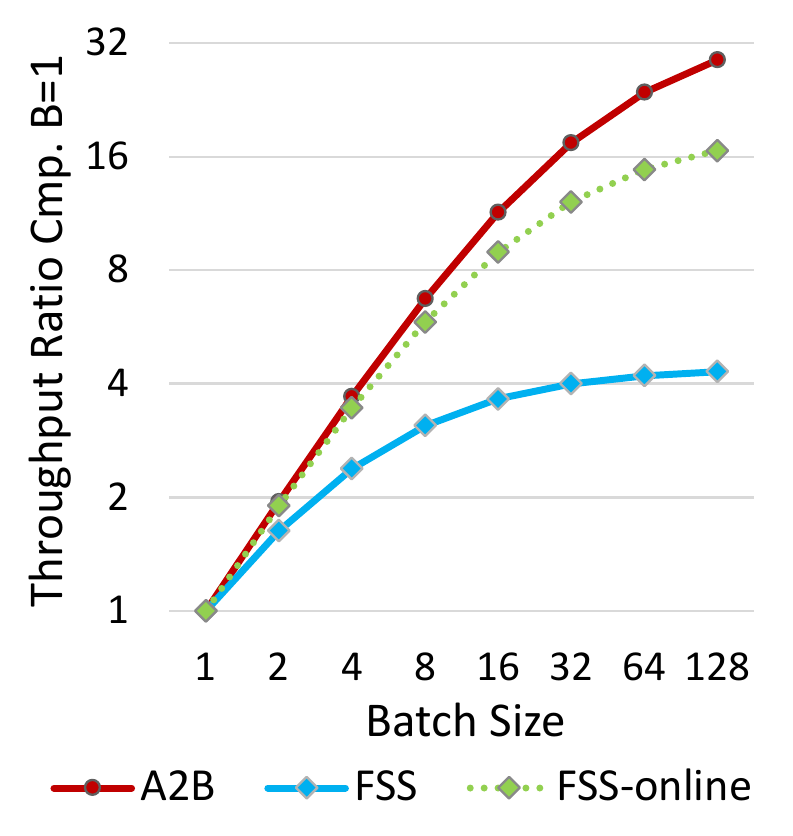}
    }
    \caption{
    Throughput (normalized to batch size 1) of \fss and \mpc with different batch sizes.
    \fwan assumed.
    }
    \label{fig:batch_perf}
    \vspace{-4mm}
\end{figure}

\textbf{Batching. }As shown in \autoref{fig:batch_perf}, batching dramatically improves MPC throughput by amortizing network latency. For \mpc under \fwan, increasing batch size from 1 to 128 improves throughput by 29$\times$ for ResNet-50 and 51$\times$ for BERT-Base. \fss benefits less with the increase of 16$\times$ and 25$\times$ (around 4.3$\times$ and 6.4$\times$ with offline included) due to its lower round complexity but high offline costs. 
The results show that using a large batch is indeed essential for MPC to achieve high throughput, as it amortizes the network round-trip latency. 
When a large enough batch size is used, MPC can be on par with or significantly outperform \fhe in terms of throughput (\autoref{tab:throughput}).

\begin{figure}[h]
    \centering
    \includegraphics[width=0.7\linewidth]{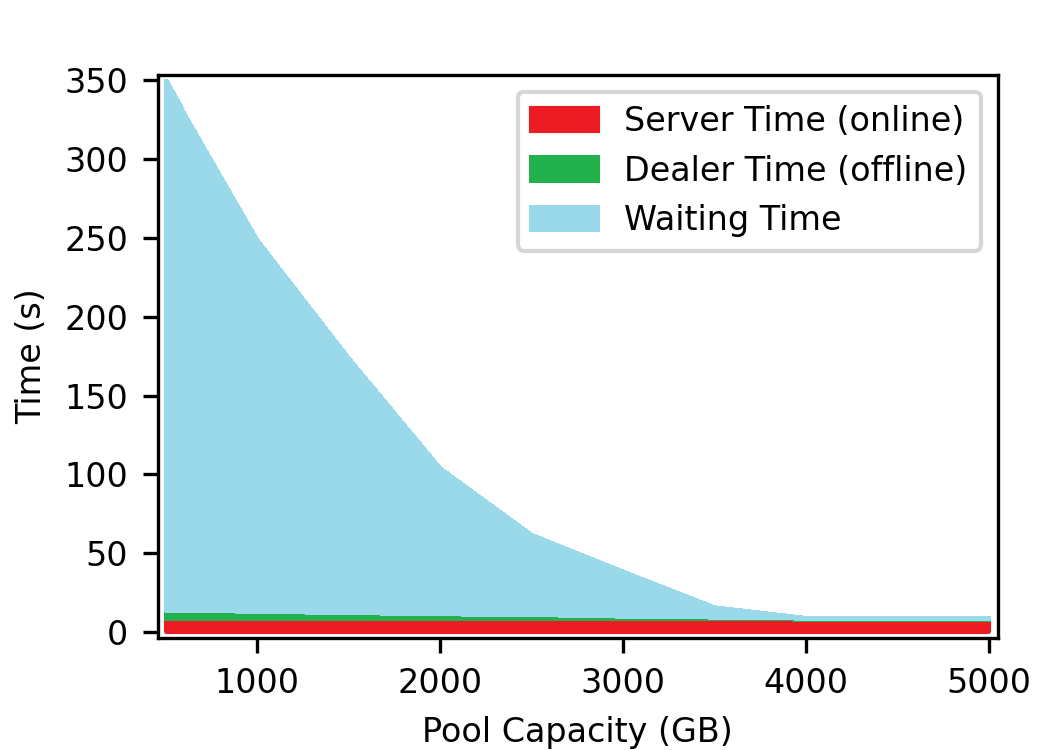}
    \caption{The \fss execution latency of the following task changes under different local key pool size. 200 jobs in \wan, each performing 128-batch inference on ResNet-20 (requires a total of around 3.8TB keys, note the performance change around the number), arriving at the server following a Poisson distribution with an average inter-arrival time of 10 seconds. 
    }
    \label{fig:processing_original}
\end{figure}

These results suggest that \fhe and \fss are better suited to scenarios where batch sizes must remain small (e.g., latency-sensitive workloads or low query arrival rates, with \fss particularly strong for online latency), whereas \fss/\mpc become preferable once larger batches are possible. As the batch size increases, the advantage of \fss in reducing communication rounds decreases, and the total volume of communication becomes the dominant factor. Batching is especially beneficial over WAN: as shown in \autoref{tab:latency}, the LAN--WAN latency gap for \mpc is enormous at batch size 1 (35--424$\times$ for \flan vs.\ \fwan), but shrinks to only 1.34--5.58$\times$ for larger batches, making MPC over WAN feasible for large-batch, throughput-oriented workloads, while LAN-like networks remain preferable for latency-sensitive use cases.


\textbf{Offline overheads of \fss.}
Although \fss is often perceived as more computation-heavy than \mpc, its offline phase is in practice dominated by communication. Because \fss relies on large pools of pre-generated keys stored locally (see \autoref{sec:mem}, \autoref{tab:mem_foot}), its performance is highly sensitive to workload patterns. The ideal online-only latency is achieved when (i) each query batch fits within the local key pool, and (ii) the long-term query rate does not exceed the combined offline+online throughput. Otherwise, queues build up, and effective throughput collapses to the much slower offline+online rate. For example, under \wan for ResNet-20, \fss throughput can drop from about 6 queries/s to 0.4 queries/s once pre-generated keys are exhausted. We also show in \autoref{fig:processing_original}, using \wan ResNet-20 as an example, that insufficient keys can cause job waiting time to spike to over 300\,s, as the parties performing the online computation have to stall while waiting for fresh keys from the dealer.

\begin{figure}[t]
    \centering
    \includegraphics[width=1\linewidth]{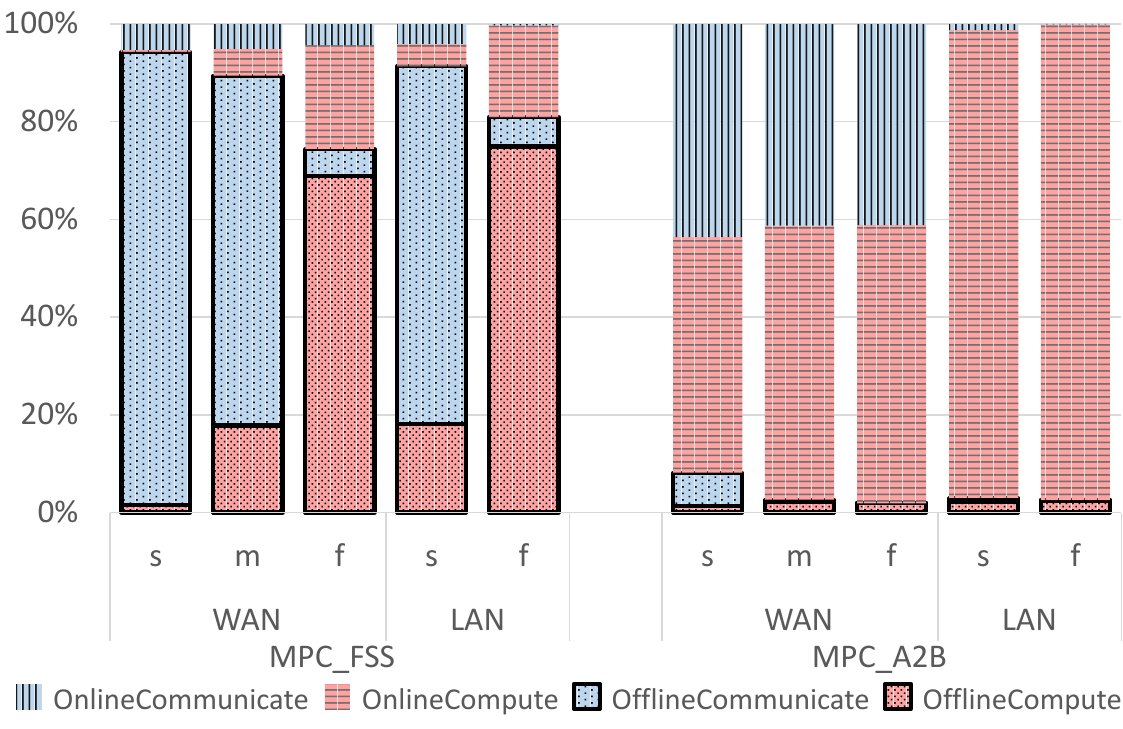}
    \caption{The proportion of total online+offline execution time of \fss and \mpc schemes under different networks for batch size 128, BERT-base.
    }
    \label{fig:time_breakdown}
\end{figure}



\autoref{fig:time_breakdown} provides a breakdown of execution time that reflects the overall trends discussed in this section.
For \fss, end-to-end performance is dominated by offline costs under slow networks and gradually shifts toward computation as network bandwidth increases.
For \mpc, online communication remains the primary bottleneck across most network settings due to its higher (amortized) round complexity, with computation becoming solely dominant in LAN environments.

In summary, \fss offers strong online performance, but incurs substantial offline cost, which can be observed in the online vs. online+offline performance gap in~\autoref{tab:latency} and~\autoref{tab:throughput}. 
\mpc can be slower online, but can have better worst-case performance than \fss in a low-performance network like \wan, 
and FHE achieves a more predictable performance dominated by computation. 

\subsection{Monetary Cost}
\begin{figure*}[t]
    \centering
    \includegraphics[width=1\linewidth]{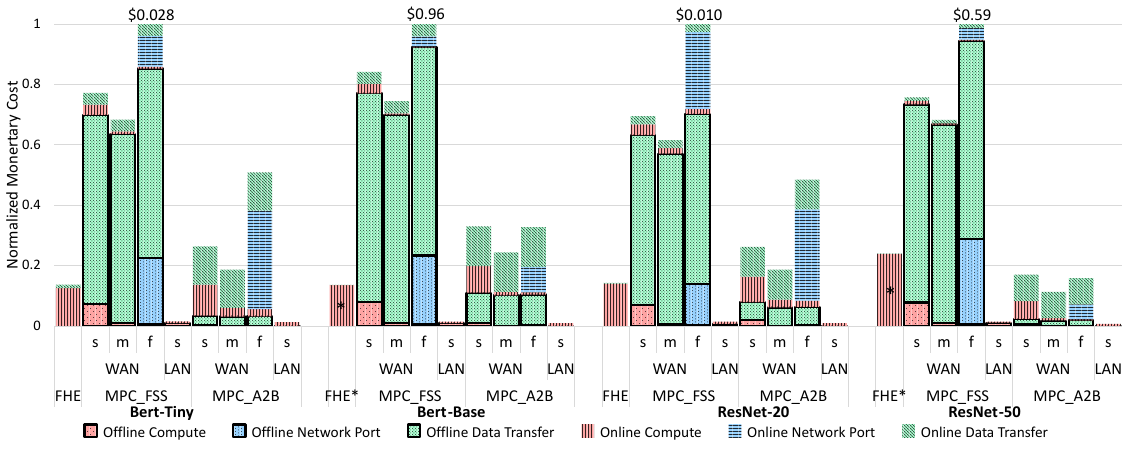}
    \caption{The total monetary cost of each inference or token generation of \fss and \mpc schemes and FHE under different settings, with batch size=128. The lengths of the bars are normalized by the largest value among the three approaches. $\ast$ denotes estimated results.}
    \label{fig:money}
\end{figure*}

The monetary costs of encrypted computation schemes are shown in~\autoref{fig:money}. The monetary cost does not scale in the same pattern as latency or throughput. For MPCs, the cost is dominated by charges for communication volume. In particular, for \fss, offline data transfer accounts for the majority of the cost (see the light-green outlined bars in \autoref{fig:money}).
Also, moving to a faster WAN can result in a higher infrastructure cost (port charges for the AWS pricing model).  
This reveals a critical system-level trade-off: \fss’s low-latency online phase relies on a communication-heavy offline phase, where a relatively high bandwidth requirements 
may dominate monetary cost at scale and cannot be easily amortized.
Thus, while \fss excels in online performance, its total cost of ownership can be high for high-throughput or data-heavy deployments.

\fhe exhibits a consistent, relatively low, and compute-dominated monetary cost profile. 
\mpc often emerges as the more cost-effective option than \fss due to less communication volume, sometimes even lower than \fhe. While \mpc and \fss protocols show cost efficiency and low latency under LAN conditions, these results must be interpreted cautiously because of the limited practicality of the LAN-based MPC threat model.

\subsection{Energy}\label{sec:energy}

\begin{figure*}[h]
    \centering
    \includegraphics[width=1\linewidth]{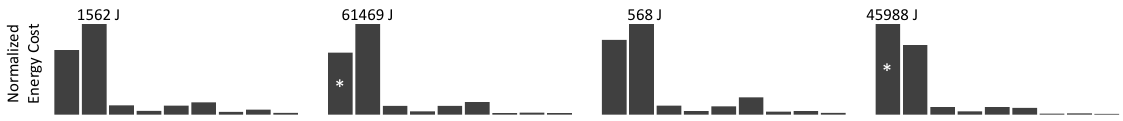}
    \includegraphics[width=1\linewidth]{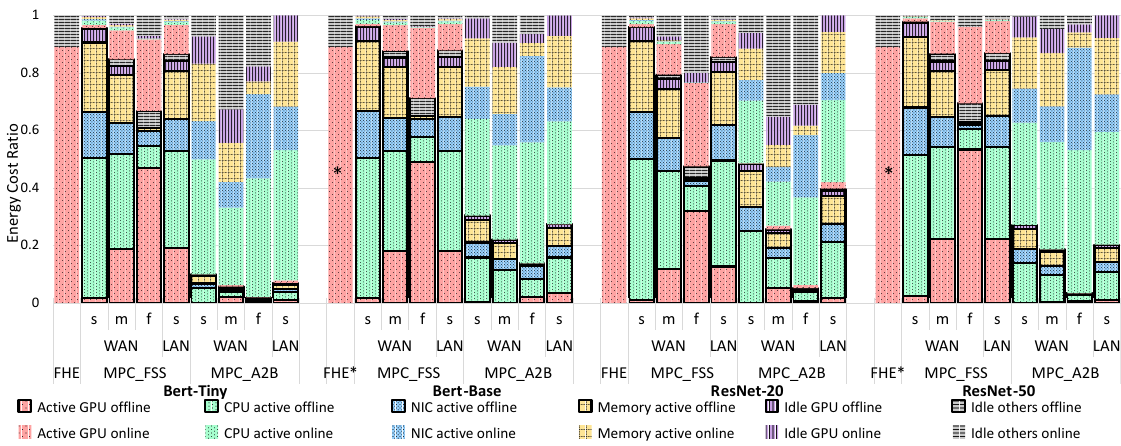}
    \caption{Total energy cost per token generated using the \fss and \mpc schemes and \fhe across different model/network settings (above, normalized by the largest value among the three approaches), along with the proportion of energy cost contributed by each method (below), with batch size=128. $\ast$ denotes estimated results.}
    \label{fig:energy_per}
\end{figure*}

We show the energy consumption of each protocol in~\autoref{fig:energy_per}. The idle cost of the whole machine, excluding the GPU idle energy, is included in the ``others'' idle energy cost. 

Across models and network settings, FHE exhibits the highest total energy consumption in most regimes other than \fss in \wan, largely due to the heavy GPU computation, which is insensitive to network configurations.
\mpc consistently incurs the lowest energy cost, while remaining sensitive to communication conditions.
\fss is dominated by heavy offline network activity, which is composed of CPU, memory, and NIC active energy.
In particular, higher performance WANs significantly reduce energy consumption for \mpc and \fss, indicating that communication-related overhead is a primary contributor to system-level power consumption. For the energy consumption of sending data through WAN, roughly 100 nJ/bit (or 850 J/GB) is estimated from previous work~\cite{aleksic2011energy,vishwanath2015energy}, which can be large but is not directly paid by the service provider, therefore not included in the figures.

\autoref{fig:energy_per} reveals that a significant portion of energy in \mpc execution is wasted as idle CPU/GPU time during WAN communication rounds (top purple and gray bars without outline), confirming that a key bottleneck is waiting for communication, not computation. 
For smaller models (Bert-Tiny, ResNet-20), up to nearly 50\% of total energy can be wasted while waiting for communication.
%
Previous work proposed adopting pipelining to effectively leverage such 
idle time during communication~\cite{wang2024mpc, harth2025pigeon}, but the benefit of pipelining is limited with slower networks (WAN)~\cite{wang2024mpc} and/or smaller batch sizes~\cite{harth2025pigeon}, as network latency becomes too large relative to computation and cannot be easily hidden.
The per-sample energy efficiency also improves with larger batches, as the impact of the slow network RTT is amortized when sending a large batch of data.
%
%
While a larger batch size improves throughput and energy efficiency, it increases response time and cannot be a solution when the application is latency-sensitive. 
The fundamental tension between latency, throughput, and resource utilization points to a non-trivial scheduling and orchestration problem for future MPC systems, particularly when optimizing for energy efficiency.

Another important observation is the substantial CPU active energy consumed by \mpc and \fss protocols during communication phases, despite low utilization. Combined with the idle CPU energy during network round-trips, this highlights a key efficiency gap; the CPU is underutilized, yet draws significant power. This slack presents an opportunity to amortize energy costs through lightweight pipelined tasks. In addition, hardware optimization on the CPU-side is as critical as GPU acceleration or network improvements to improve energy efficiency \mpc / \fss. 



\subsection{Memory footprint}\label{sec:mem}
~\autoref{tab:mem_foot} reports the measured CPU and GPU memory usage for all schemes. \fhe uses minimal CPU memory, but several GPU entries are missing due to out-of-memory (OOM) failures. The bracketed values denote the theoretical maximum usage when the layers are loaded independently. Although NEXUS~\cite{zhang2024secure} reports ResNet-50 results in multi-GPU settings, their single-GPU variants cannot host the largest layers and result in OOM in our unified single-GPU environment.

\fhe is GPU memory-intensive because the ciphertext can be thousands of times larger than the plaintext scalars and requires substantial auxiliary keys (rotation, relinearization, bootstrapping), which themselves consume several GBs~\cite{reagen2021cheetah, kim2024cheddar, zhang2024secure}. Storing encrypted activations, temporary intermediates, and helper keys quickly pushes GPU usage into the tens of gigabytes. Between the two implementations included, Cheddar consumes significantly more memory compared to NEXUS, as it (i) uses a pooled GPU allocator (RMM), (ii) caches large NTT/mod-switch tables and scratch buffers, and (iii) loads many rotation and relinearization keys at startup. FHE protocols are often assumed to run on high-end, large-memory GPUs that can host all intermediate results, such as those used in prior work~\cite{reagen2021cheetah, kim2024cheddar, zhang2024secure}. However, once a layer exceeds GPU memory capacity, ciphertexts must spill into CPU memory over PCIe, incurring severe overhead. For example, offloading the largest MatMul layer can potentially result in a 30× slowdown for BERT-Base due to data transfer. This overhead grows at least linearly with larger model dimensions or longer context lengths, making spill-and-reload execution impractical for scalable deployment.

It is also important to note that \fss likewise has substantial key-storage overhead. As we have already shown in the offline communication cost, the pre-generated cryptographic keys (and Beaver triples) are huge and scale with the function size, model dimension, and total query volume. 
Previous \fss work as well as the results in our experiment section assume that the keys are loaded and stored in CPU memory before the online phase starts. If reading from the disk is needed (e.g., there is a large time gap between the offline and online phases, which makes storing keys only in DRAM impractical), read speed becomes an extra parameter affecting performance. If we assume a low 1GBps total read bandwidth from disk for BERT-Base under \mwan, this introduces 47\% more online latency and 58\% more online+offline latency for a batch size of 1, and causes throughput to drop to only 12\% of online throughput and 42\% of online+offline throughput, making \fss less appealing than \mpc even for online throughput. However, a multi-NVMe 10GB/s-class setup can restore near-ideal pipelining of disk reads and effectively eliminate this read overhead.



\begin{table}[t]
\centering
\footnotesize
\setlength{\tabcolsep}{2.56pt}
\caption{Peak memory footprint (in GB) of MPC and FHE protocols. Values in parentheses denote extrapolated usage for the layer consuming the largest GPU memory, assuming no preload or caches of other layer's data.}
\begin{tabular}{lcccccccc}
\hline
 & \multicolumn{2}{c}{BERT-T} & \multicolumn{2}{c}{BERT-B} 
 & \multicolumn{2}{c}{ResN-20} & \multicolumn{2}{c}{ResN-50} \\
Method & CPU & GPU & CPU & GPU & CPU & GPU & CPU & GPU \\
\hline
\mpc & 0.65 & 0.1 & 0.71 & 4   & 0.49 & 0.9 & 0.53 & 2   \\
\fss     & 4.2 & 0.4 & 18.9 & 1.8 & 3.6  & 0.3 & 11.7 & 2.5 \\
{\fss}(SSD) & 0.3  & --  & 16.8 & --  & 0.2  & --  & 9.9  & --  \\
\fhe     & --   & 18.6(18) & -- & --(112) & -- & 28 (4.66) & -- & --(228) \\
\hline
\end{tabular}

\label{tab:mem_foot}
\end{table}



\subsection{Scalability}

We additionally study how each approach scales with different (i) input context lengths for language inputs and (ii) more powerful future hardware setups.

\begin{figure}
    \centering
    \includegraphics[width=\linewidth]{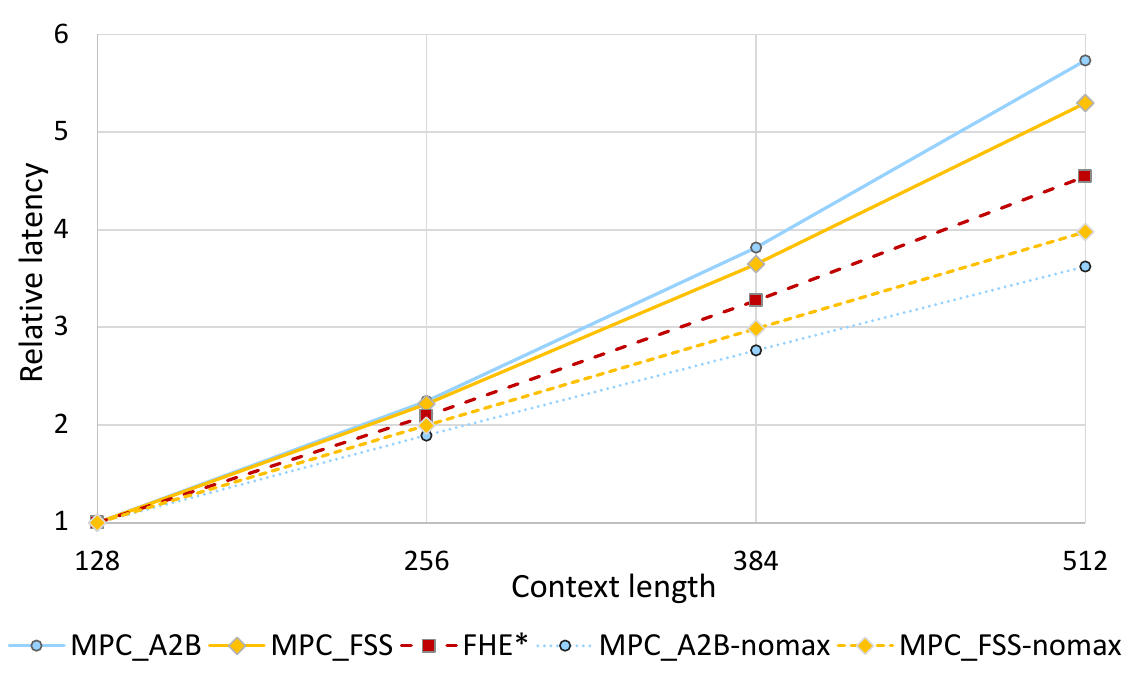}
    \caption{Relative latency change with batch size 128 and \mwan network for BERT-Base, normalized at the context length 128 result of each scheme. Numbers for layers of \fhe that would cause OOM are estimated with $\ast$ denoting estimated results.}
    \label{fig:contextlength}
    \vspace{-4mm}
\end{figure}

\textbf{Context length. }
We study how the inference latency scales with the input context length \(n\) for BERT-Base in~\autoref{fig:contextlength}. 
While self-attention incurs \(O(n^2)\) complexity for all schemes, the observed scaling differs because each approach is bottlenecked by different system costs.
Our MPC implementations include a max operation before Softmax to stabilize the approximation~\cite{crypten2020,gupta2023sigma}, whereas the FHE implementation~\cite{zhang2024secure} does not. 
For fair comparison, we additionally report the results for an MPC variant without the max (``-nomax'').

Without the max, attention is dominated by the matrix multiplication computing attention scores.
In this setting, FHE scales the worst as \(n\) increases because it is computation-bound: encrypted matrix multiplications and bootstrapping costs grow rapidly with attention size, with linear layers scaling as \(O(n)\) and attention-related computation scaling as \(O(n^2)\).
In contrast, MPC-based schemes (\mpc and \fss) are communication-bound and their overhead grows as \(O(n)\) for most layers; although attention communication grows as \(O(n^2)\), for the evaluated context lengths (\(n \leq 512\)) this cost is dominated by other layers and round-trip latency, resulting in flatter scaling. This leads to the overhead growth of FHE \(>\) \fss-nomax \(>\) \mpc-nomax.

Including the max operation introduces an additional \(O(\log n)\) round complexity and extra \(O(n^2)\) communication, which increases round-trip latency.
Because \mpc implements the max using A2B comparison protocols, it incurs higher overhead growth compared to \fss.
Consequently, with the max included, \mpc scales worse than \fss.

Beyond latency, the schemes also differ in hardware resource requirements as context length increases.
FHE suffers from at least linearly increasing GPU memory usage due to ciphertext expansion and intermediate buffers, which limits feasible context lengths and causes out-of-memory failures for larger models.
MPC-based schemes require less GPU memory but incur growing communication costs, and \fss additionally requires at least linearly increasing CPU memory and disk space for FSS key material.
These results highlight the need for co-designing PPML schemes, model architectures, and system resources for long-context Transformer inference.

\begin{figure*}[h]
    \centering
    \subfloat[BERT-Base Online]{%
        \includegraphics[width=0.24\textwidth]{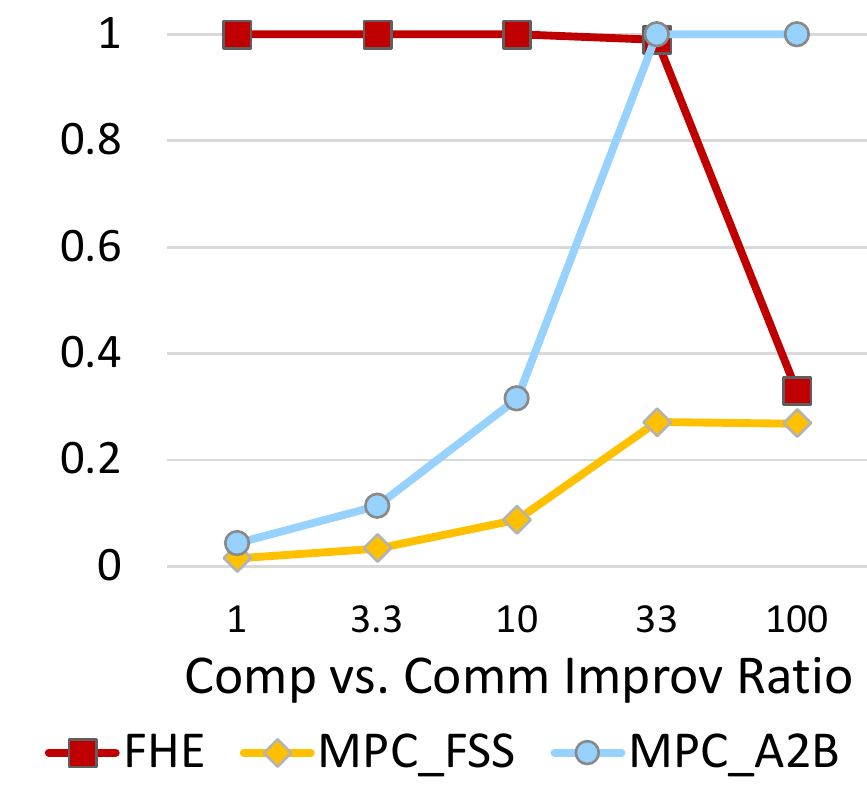}%
        \label{fig:sub1}%
    }\hfill
    \subfloat[BERT-Base Online+Offline]{%
        \includegraphics[width=0.24\textwidth]{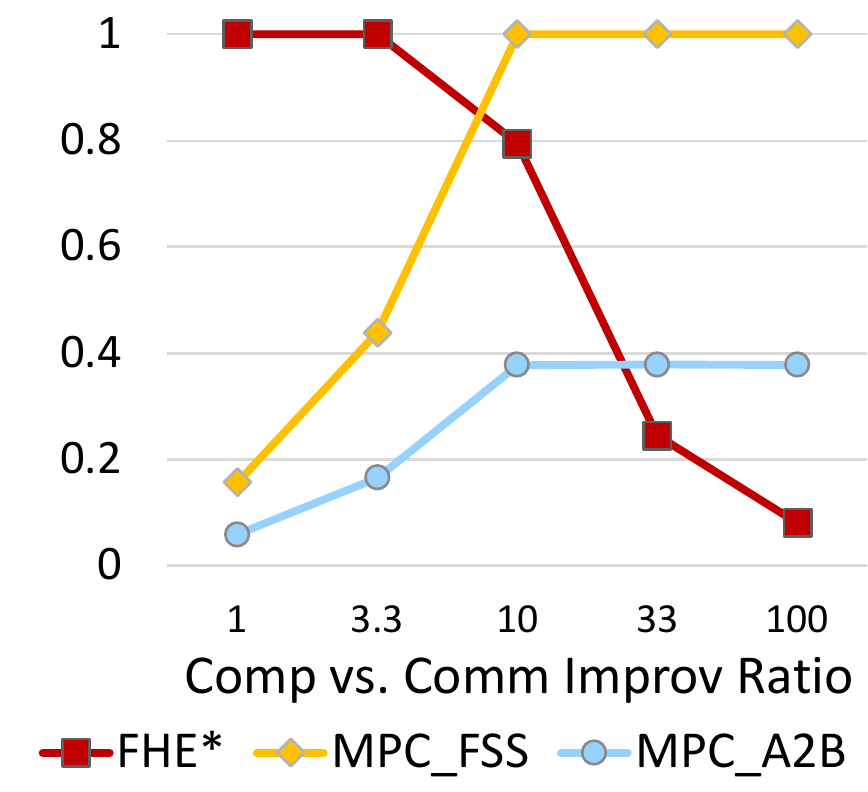}%
        \label{fig:sub2}%
    }\hfill
    \subfloat[ResNet-50 Online]{%
        \includegraphics[width=0.24\textwidth]{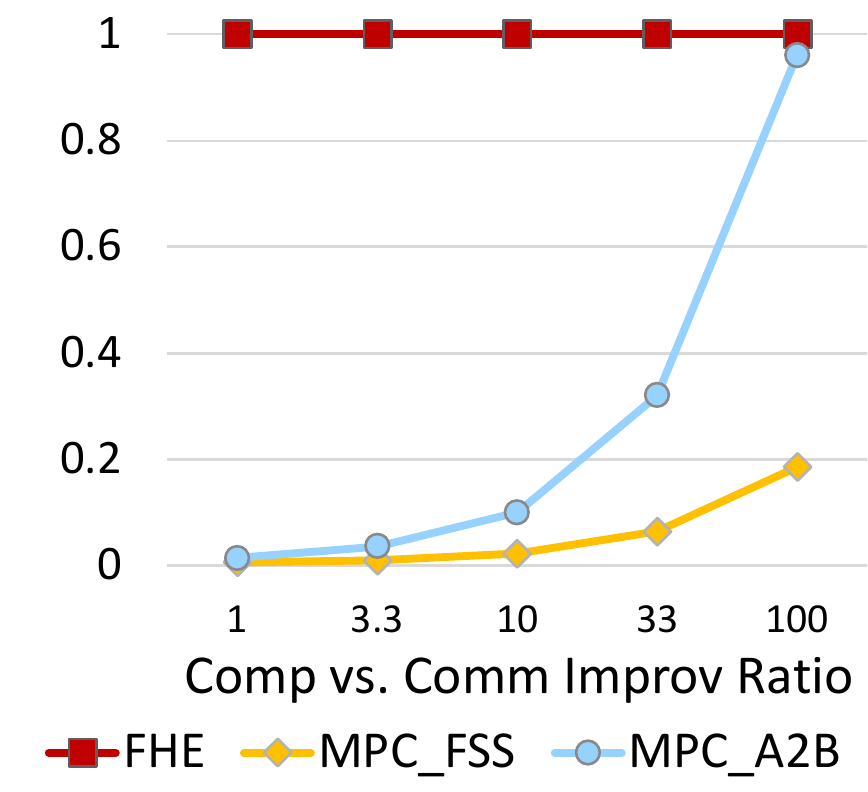}%
        \label{fig:sub3}%
    }\hfill
    \subfloat[ResNet-50 Online+Offline]{%
        \includegraphics[width=0.24\textwidth]{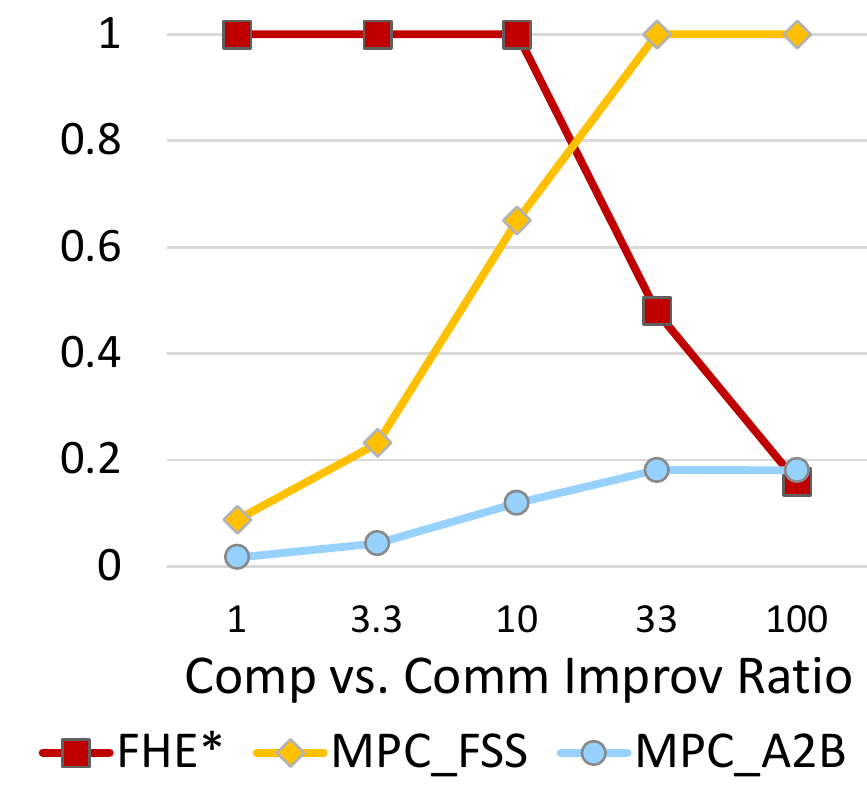}%
        \label{fig:sub4}%
    }
    \caption{The relative latency of \mpc, \fss and FHE when the computation hardware improves faster than the communication hardware, based on batch size 128 \mwan results. The x-axis represents how many times faster computation becomes relative to communication compared to the baseline system in our previous experiment. 
    The y-axis shows normalized latency across protocols as hardware capabilities evolve under this model. $\ast$ denotes estimated results.}
    \label{fig:scaling}
\end{figure*}

\textbf{Hardware advancement.} Our earlier results compare the performance of FHE and MPC schemes on our current hardware setup with a set of network configurations. \new{Our setup reflects moderate, not highest-end hardware,} and the relative performance may change with higher-performance hardware, now and in the future; as FHE is mostly compute-bound, \mpc is communication-bound, and \fss is in-between, they will be affected differently by hardware and network performance improvements. 

Previous work indicated that network latency, especially wide-area RTT, has improved only slowly~\cite{rumble2011s}, largely due to fundamental propagation limits and the economic cost of deploying faster long-haul infrastructure. Computing hardware has historically advanced at a much faster pace, driven by improvements in semiconductor technology, architectural specialization, and domain-specific accelerators~\cite{hennessy2019new,theis2017end}.
Together, these observations suggest that computing throughput is likely to continue to improve more rapidly than WAN communication latency/bandwidth, making \fhe relatively more competitive.

%
\autoref{fig:scaling} further explores how the relative performance of the three schemes changes as the computation (CPU/GPU) and communication (network) improve at different rates.  
The x-axis shows the ratio of improvement between computation and communication bandwidth, e.g., $x=100$ means that computation improves 100$\times$ more compared to communication.
The y-axis shows the normalized latency, normalized by the slowest of the three approaches.

%
As computation becomes relatively faster, both
\mpc and \fss improve in absolute terms, but their \emph{relative} latency increases 
because \fhe benefits much more from increased computational capability. A notable observation is
that even under a 100$\times$ compute–communication imbalance, \fhe still struggles to match the
online performance of \fss. Although \fhe accelerates quickly with compute-heavy
scaling, the online phase of \fss is so lightweight that it maintains a strong normalized
advantage while also benefiting from the growth of computing power. 
\new{A recent concurrent work~\cite{jayashankar2025scalable} reporting BERT-Base FHE on A100/H100/B200 GPUs aligns with our results, as even 8 B200 GPUs in their work achieve lower throughput than \fss+\fwan in our setup.}
However, in the online+offline setting, the substantial offline communication footprint
of \fss limits its relative gains, causing its latency to quickly become the slowest. Overall, these results show that compute-centric hardware advances shift the balance in
favor of \fhe, but MPC-based schemes---and in particular the fast online phase of \fss---retain an advantage when offline preprocessing can be hidden with sufficient resources. It is also worth noting that ResNet, with heavy linear computation, favors \fhe more than BERT, which has more complex non-linear operations.



\subsection{Key takeaways}\label{sec:takeaways}
We summarize our key findings below.

\textbf{Latency and throughput.}
Batching is essential for MPC for high throughput and energy efficiency. With a large-enough batch, the total communication volume, not the number of communication rounds, becomes the primary scalability bottleneck for both \mpc and \fss. The FHE latency is dominated by local computation and remains insensitive to network performance.

\textbf{Offline cost.}
Offline overhead, especially for \fss, is significant in terms of both energy and performance.
Unlike prior work's assumptions, the latency of the offline phase cannot always be hidden when the query arrival rate exceeds the offline phase throughput, and the energy/monetary cost of the offline phase cannot be hidden at all.

\textbf{Energy cost.}
Our study shows that CPU/GPU idle energy during communication accounts for a non-negligible portion in MPC, calling for more research on protocol-level optimizations or parallel scheduling of jobs to minimize these idle cycles. Our study also exposes the importance of optimizing CPU energy efficiency for MPC.

\textbf{Monetary cost.}
The trend of monetary cost does not necessarily follow that of latency/throughput. The monetary cost shows less dependency on the network bandwidth for MPC but more on the total communication volume. FHE avoids the cost of WAN bandwidth but pays for intensive GPU computation.

\input{dem}

For the above cost metrics, we show a summary of their trade-off in~\autoref{fig:dem}.

\textbf{Memory footprint.}
\fss requires hundreds to thousands of GB of offline key storage in disks, \fhe requires tens of GB of GPU memory during inference, while \mpc can run with lower memory usage. These asymmetries fundamentally shape which schemes are deployable for large models or long-context workloads with limited resources.

\textbf{Scalability.}
\fhe scales worse than MPC in our evaluated regimes with increasing model dimension or input context length. Computation hardware improvements can make \fhe competitive or superior for high-volume workloads, but \fss remains more efficient for online latency when network bandwidth and storage are provisioned. We also highlight the necessity of co-designing PPML protocols and ML model architectures to meet future scalability demands.

\section{Conclusions}
Our characterization reveals new system-level insights for PPML protocols.
We show that PPML trade-offs cannot be understood through online latency alone. Offline overheads, communication volume, idle energy, and memory footprint fundamentally shape deployability and cost, and conclusions drawn from latency-only evaluations can therefore be misleading. These findings highlight the need for system-aware evaluation and joint consideration across PPML protocols, ML model architectures, and hardware platforms to enable practical large-scale private inference.

\section*{Acknowledgments}
This work is partly supported by the U.S. National Science Foundation under award No. CCF-2118709,  CNS-2349610 and CCF-2529883. Any opinions, findings, and conclusions or recommendations expressed in this material are those of the author(s) and do not necessarily reflect the views of the National Science Foundation.

\bibliographystyle{IEEEtran}
\bibliography{BIBFILE.bib}

\newpage

\end{document}

%% file: dem.tex
\begin{figure}[t]
    \centering
    \fontsize{8pt}{9.6pt}\selectfont
    \subfloat[\parbox{3cm}{\centering Scheme with best\\throughput/latency}\label{fig:latency_sub}]{
    \begin{minipage}[b]{0.47\linewidth}
        \begin{tikzpicture}[scale=0.45,>=latex]

          \coordinate (O)  at (0,0);   
          \coordinate (X)  at (7,0);   
          \coordinate (Y)  at (0,7);   
          \coordinate (H)  at (7,3); 
          \coordinate (J)  at (2.5,2.6); 
          \coordinate (JT) at (3,7);   
          \coordinate (XR) at (7,7);   
        
        
          \fill[blue!25]
            (O) -- (Y) -- (JT) -- (J) -- cycle;
        
          \fill[red!25]
            (O) -- (X) -- (H) -- (J) -- cycle;
        
          \fill[green!25]
            (J) -- (JT) -- (XR) -- (H) -- cycle;
        
          \draw[very thick,->] (O) -- (X);
          \draw[very thick,->] (O) -- (Y);
        
          \node[below] at ($(O)!0.5!(X)$) {\selectfont Communication};
          \node[rotate=90,above] at ($(O)!0.5!(Y)$) {\selectfont Computation};
        
          \draw[very thick] (O) -- (J);
          \draw[very thick] (J) -- (JT);
          \draw[very thick] (J) -- (H);
        
          \node at (1.2,4.2) {\selectfont FHE};
          \node at (4,1.3) {\selectfont MPC A2B};
          \node at (5,4.8) {\selectfont MPC FSS};
        
        \end{tikzpicture}
        \vspace{0pt}
        \end{minipage}
    }
    \subfloat[\parbox{3cm}{\centering Energy/Power \\relative cost
    }\label{fig:throughput_sub}]{
    \begin{minipage}[b]{0.47\linewidth}
    \footnotesize
    \setlength{\tabcolsep}{2.8pt}
        \begin{tabular}{l c c}
        
        \multicolumn{3}{c}{\textbf{Monetary Cost}} \\
        \hline\hline
        FHE & \multicolumn{2}{c}{Comp.} \\\hline
        \multirow{2}{*}{\mpc} & Online & Data \\
                             & Comp. & Transfer \\\hline
        \multirow{2}{*}{\fss} & Offline Data & \multirow{2}{*}{Comp.} \\
                             & Transfer     &   \\
        \hline\hline
        \multicolumn{3}{c}{\textbf{Energy Cost}} \\
        \hline
        FHE & \multicolumn{2}{c}{Comp.} \\\hline
        \mpc & Idle & Comm. \\\hline
        \fss & Comp. & Comm.  \\\hline
        \hline
        \end{tabular}
        \vspace{0pt}

        \end{minipage}

    }
    \caption{Summary of cost metrics trade-offs. (a) illustrates the relative regimes in which each scheme is advantageous for throughput and latency as communication capability (e.g., higher network bandwidth or lower RTT) and computation capability (e.g., CPU/GPU performance) improve; (b) shows 2 main components of the monetary and energy costs for each scheme.}
    \label{fig:dem}
    \vspace{-2mm}
\end{figure}
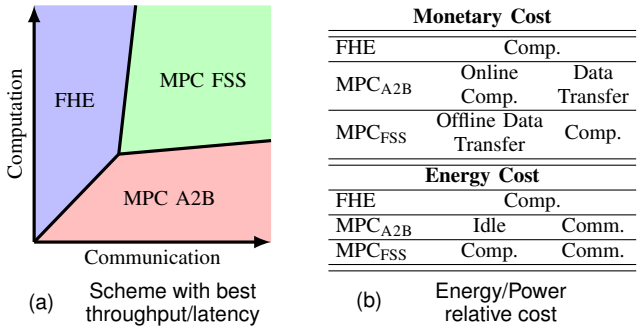